\documentclass[published]{JHEP}

\usepackage{amssymb}

\newcommand{\Sl}{\mathop{\rm Sl}\nolimits}
\newcommand{\SO}{\mathop{\rm SO}\nolimits}
\newcommand{\STr}{\mathop{\rm STr}\nolimits}
\newcommand{\Sp}{\mathop{\rm Sp}\nolimits}
\newcommand{\U}{\mathop{\rm {}U}\nolimits}

\newcommand{\irrep}[1]{\mathbf{#1}}
\newcommand{\eis}[3]{{\cal E}^{#1}_{\irrep{#2};\,#3}}
\newcommand{\exc}{E_{d+1(d+1)}}
\newcommand{\vect}{\mathbf{V}}
\newcommand{\Zint}{\mathbb{Z}}
\newcommand{\Real}{\mathbb{R}}
\newcommand{\M}{{\cal M}}
\newcommand{\F}{{\cal F}}
\newcommand{\zclass}{\mathcal{A}_{\mathrm{BPS}}}

\title{$R^4$ couplings, the fundamental membrane\\
and exceptional theta correspondences}

\author{Boris Pioline\thanks{On
leave of absence from LPTHE, Universit{\'e} Pierre et Marie Curie,
PARIS VI and Universit{\'e} Denis Diderot, PARIS VII, Bo\^{\i}te
126, Tour 16, 1$^{\it er}$ {\'e}tage, 4 place Jussieu, F-75252
Paris CEDEX 05, FRANCE}\\
Jefferson Physical Laboratory, Harvard University,
Cambridge, MA 02138, USA\\
Email: \email{pioline@physics.harvard.edu}}

\author{Hermann Nicolai and Jan Plefka\\
Albert-Einstein-Institut, Max-Planck-Institut
f\"ur Gravitationsphysik\\Am M\"uhlenberg 1, D-14476 Golm, Germany\\
Email: \email{nicolai,plefka@aei-potsdam.mpg.de}}

\author{Andrew Waldron\thanks{On leave of absence from Dept. of
Mathematics, UC Davis, CA 95616.}\\
Physics Department, Brandeis University, Waltham, MA 02454, USA\\
Email: \email{wally@brandeis.edu}}

\abstract{This letter is an attempt to carry out a first-principle
computation in M-theory using the point of view that the
eleven-dimensional membrane gives the fundamental degrees of
freedom of M-theory. Our aim is to derive the exact BPS $R^4$
couplings in M-theory compactified on a torus $\mathbb{T}^{d+1}$
from the toroidal BPS membrane, by pursuing the analogy with the
one-loop string theory computation. We exhibit an $\Sl(3,\Zint)$
modular invariance hidden in the light-cone gauge (but obvious in
the Polyakov approach), and recover the correct classical spectrum
and membrane instantons; the summation measure however is
incorrect. It is argued that the correct membrane amplitude
should be given by an exceptional theta correspondence lifting
$\Sl(3,\Zint)$ modular forms to $\exc(\Zint)$ automorphic forms,
generalizing the usual theta lift between $\Sl(2,\Zint)$ and
$\SO(d,d,\Zint)$ in string theory. The exceptional correspondence
$\Sl(3)\times E_{6(6)}\subset E_{8(8)}$ offers the interesting
prospect of solving the membrane small volume divergence and
unifying membranes with five-branes.}

\keywords{M-Theory, String Duality}

\received{February 22, 2001}

\accepted{March 23, 2001\smash{\llap{\raise-1\baselineskip
\hbox{{\sc Revised:} March 22, 2001}}}}

\JHEP{03(2001)036}

\begin{document}

\paragraph{Introduction.}

\looseness=1 Despite considerable insights brought about by the
discovery of dualities, a tractable microscopic definition of
non-perturbative string theory remains an outstanding problem
hindering further progress. While several definitions are
available, computational power is the issue: the
eleven-dimensional membrane is strongly interacting, whereas the
large-$N$ limit of its M(atrix) theory cousins is still largely
untamed. For  particular supersymmetric
situations however, one would expect all quantum fluctuations to
cancel, leaving a hopefully tractable zero-mode problem. For
example, the spectrum of BPS states in toroidal compactifications
of M-theory has been reproduced from the BPS
five-brane~\cite{Dijkgraaf:1997hk} or from M(atrix) theory (see
e.g.~\cite{Obers:1998fb} for a review along these lines). This
amounts to a rather trivial check of the duality invariance of
the proposed classical action. Somewhat less trivial would be to
give a microscopic derivation of a particular amplitude,
receiving contributions from supersymmetric states but still with
a non-trivial summation measure. Such is the case of the
eight-derivative $R^4$ couplings, exactly known in toroidal
compactifications of M-theory on the basis of duality and
supersymmetry~\cite{Green:1997tv}--\cite{Obers:1999um}. In this
work, we attempt a microscopic derivation of these couplings,
contending that the eleven-dimensional membrane gives the
fundamental degrees of freedom of M-theory. Due to our limited
understanding of the quantization of the membrane, we proceed by
analogy with the BPS string, and try to construct a partition
function for a toroidal BPS membrane exhibiting at the same time
world-volume modular invariance and target-space
U-duality.\footnote{Our goal bears some resemblance to the
work~\cite{nappi}, where a modular invariant partition function
for the fivebrane is proposed; unfortunately their result is not
invariant under target-space duality.} The classical membrane
partition function reproduces the correct BPS spectrum and
instanton saddle points, but is not U-duality invariant. We
propose that the appropriate quantum modification is provided by
exceptional theta correspondences from the theory of automorphic
forms, in analogy with the symplectic theta correspondence
arising in string theory. The explicit construction of such
correspondences is outlined but left for future work. It is
nevertheless expected to provide important clues about the
quantization of the membrane, and possibly give a mechanism for
the finiteness of membrane theory and unification of membranes
and five-branes.

\paragraph{One-loop string amplitude.}

In order to motivate our approach, recall the computation of $R^4$
couplings in type-II string
theory~\cite{Green:1997tv,Kiritsis:1997em}. The four-graviton
scattering amplitude at eight derivative order receives
contributions from tree-level, one-loop, together with an
infinite series of D-instanton corrections,
\begin{equation}\label{pertexp}
f_{R^4}= \frac{2\zeta(3)V}{g_s^2 l_s^2} +
f_{1\hbox{-}\mathrm{loop}} + \mathcal{O}\left(e^{-1/g_s}\right).
\end{equation}
The tree-level term is essentially field-theoretical and depends
only on the total volume $V$ of the internal torus
$\mathbb{T}^d$. The one-loop term on the other hand is a
complicated function of the torus moduli, invariant under target
space duality $\SO(d,d,\Zint)$. It can be written as an integral
over the fundamental domain of the Poincar\'e upper half-plane,
\begin{equation}\label{1loop}
f_{1\hbox{-}\mathrm{loop}}= 2\pi \int_\F \frac{d^2\tau}{\tau_2^2}
Z_{d,d}\, ,\qquad (\tau;g,B)\,,
\end{equation}
where $Z_{d,d}$ is the partition function (or theta function) of
the even self-dual lattice describing the toroidal
compactification,
\begin{equation}\label{lag}
Z_{d,d}(\tau;g,B) = V \sum_{(m^i,n^i)\in\Zint^{2d}}
e^{ -\frac{\pi}{\tau_2} (m^i + \tau n^i ) g_{ij}(m^j + \bar \tau
n^j) +2\pi i m^i B_{ij} n^j}\, .
\end{equation}
Notice that only the zero-modes of the string coordinates
contribute, all bosonic and fermionic oscillators having canceled
in this supersymmetric amplitude. Their weight is given by the
Polyakov action
\begin{equation}
\label{pols}
S=\int d^2 \sigma \sqrt{\gamma} g_{ij} \gamma^{ab}
\partial_a X^i\partial_b X^j + 2\pi i \epsilon^{ab} B_{ij}
\partial_a X^i\partial_b X^j \, ,
\end{equation}
evaluated on the classical (zero-mode) configuration and
worldsheet unit-volume \mbox{metric}
\begin{equation}
X^i= m^i \sigma_1 + n^i \sigma_2\, ,\qquad \gamma_{ab}=
\frac{1}{\tau_2}\pmatrix{1&\tau_1\cr \tau_1&|\tau|^2}.
\end{equation}

\paragraph{Sum over states.}

For the purpose of generalization to the eleven-dimensional
membrane, it will be useful to understand this one-loop amplitude
from several viewpoints. First, let us Poisson resum the windings
$m^i$ along the $\sigma_1$ direction into momenta $m_i$, and
rewrite
\begin{equation}
\label{ham} f_{1\hbox{-}\mathrm{loop}} =2\pi \int_\F
\frac{d^2\tau}{\tau_2^{2-(d/2)}} \sum_{m_i,n^i} e^{ -  \pi \tau_2
\left[ (m_i+B_{ij}n^j)^2 + (n^i)^2 \right] -2\pi \tau_1 i m_i
n^i}\, .
\end{equation}
Ignoring for a moment the restriction $|\tau|>1$ on the
fundamental domain $\F$, we recognize $\tau_2$ as the Schwinger
parameter, while the integral over $\tau_1\in [-1/2,1/2]$ imposes
the BPS constraint $m_i n^i=0$. The one-loop amplitude can
therefore be viewed as a \emph{regulated} sum over the one-loop
contributions of all perturbative half-BPS states, with internal
momentum $m_i$, winding $n^i$ and mass
\begin{equation}
\M^2= \left(m_i+B_{ij}n^j\right)^2 + \left(n^i\right)^2 .
\end{equation}
In fact, one could have started from the light-cone description of
the string to find the integrand in~(\ref{ham}), discovered its
hidden modular invariance under $\Sl(2,\Zint)$, and by hand
restricted the integration to the fundamental domain, to render
the integral UV finite. The Polyakov description produces this
result automatically, and shows that all string theory amplitudes
are UV finite.

\paragraph{Worldsheet instantons.}

Let us now revert to the ``lagrangian''
representation~(\ref{lag}), and consider the large volume
expansion (or equivalently the Fourier expansion in the periodic
modulus $B_{ij}$) of the one-loop result. By the standard orbit
decomposition method~\cite{Dixon:1991pc}, the fundamental domain
can be unfolded to produce
\begin{eqnarray}
f_{1\hbox{-}\mathrm{loop}}&=&\frac{2\pi^2}{3}V +2V \sum_{m^i\neq
0}\frac{1}{ m^{i}g_{ij}m^{j}}+\nonumber\\ &&+\,4\pi V
\sum_{m^{ij}/\Sl(2)} \mu\left(m^{ij}\right) \frac{\exp[-2\pi
\sqrt{(m^{ij})^2}+2\pi i B_{ij} m^{ij}] }{\sqrt{(m^{ij})^2}}
\label{largev}
\end{eqnarray}
corresponding to the rank 0, 1 and 2 orbits of the integers
$(m^i,n^i)$ under the modular group $\Sl(2,\Zint)$, respectively.
Again, the first and second term can be interpreted as the
regulated contribution from the Kaluza-Klein states propagating
in the loop, while the last term corresponds to worldsheet
instantons wrapping subtori $\mathbb{T}^2\subset \mathbb{T}^d$
with homology $m^{ij}=m^i n^j-m^j n^i$. Their classical action is
recognized as the Nambu-Goto action, which follows from the fact
that the Polyakov and Nambu-Goto action are equivalent upon
imposing the equations of motion for the worldsheet metric
$\gamma_{ab}$. More interestingly, the summation measure can be
computed as the number theoretic~function
\begin{equation}\label{mes}
\mu\left(m^{ij}\right)= \sum_{n|m^{ij}} n
\end{equation}
which is also fixed by requiring T-duality invariance. The
measure~(\ref{mes}) follows from the fact that the sum over the
integers $(m^i,n^i)$ is restricted to $\Sl(2,\Zint)$ orbits,
which can be parameterized as
\begin{equation}
\pmatrix{m^i\cr n^i}= \pmatrix{ m &j  & m^3 & m^4 &\cdots \cr 0 &
n & n^3 & n^4 &\cdots}
\end{equation}
with $m,n>0$ and $0\leq j< n$. The factor $n$ in~(\ref{mes})
arises from summing over $j=0,\dots, n-1$. Clearly, this factor
would not be accessible from a Nambu-Goto description of the
string.

\paragraph{Automorphic forms and theta lift.}

Finally, the one-loop result~(\ref{largev}) can be encapsulated
into a manifestly T-duality invariant result, using Eisenstein
series of the T-duality group~\cite{Obers:1999um}. A particularly
convenient representation is in terms of the vector
representation,\footnote{More precisely, the one-loop amplitude is
equal to the \emph{residue} of the Eisenstein series at the pole
at $s=d/2-1$. We thank D.~Kazhdan for pointing this out to us.}
\begin{equation}\label{c2}
f_{1\hbox{-}\mathrm{loop}}= 2
\frac{\Gamma({d}/{2}-1)}{\pi^{{d}/{2}-2}}
\eis{\SO(d,d,\Zint)}{\vect}{s={d}/{2}-1}\,,
\end{equation}
where the divergence is regulated by analytic continuation to the
complex $s$-plane, as in standard zeta function regularisation.
This identification is in particular supported by the
identity~\cite{Obers:1999um}
\begin{equation}
\label{lapl}
\left[\Delta_{\SO(d,d)}-2 \Delta_{\Sl(2)}
+\frac{d(d-2)}{4}\right] Z_{d,d} = 0
\end{equation}
satisfied by the BPS string partition function: after integrating
by parts, this implies that the one-loop contribution is an
eigenmode of the invariant laplacian on
$\SO(d,d,\Real)/\SO(d)\times \SO(d)$ with eigenvalue $d(d-2)/4$,
as is the case of the r.h.s.\ in~(\ref{c2}). In fact, $Z_{d,d}$
and $f_{1\hbox{-}\mathrm{loop}}$ are eigenmodes of all invariant
differential operators on $\Sl(2)/\U(1) \times
[\SO(d,d,\Real)/\SO(d)\times \SO(d)]$.

\looseness=1 For higher point functions, the structure of the
one-loop amplitude remains essentially as in~(\ref{1loop}),
except for the insertion of a modular form $\Phi(\tau,\bar\tau)$
incorporating the effect of the oscillators and vertex
insertions. The integrated amplitude is still T-duality
invariant. From the mathematical point of view, this provides a
``theta'' lift of $\Sl(2,\Zint)$ modular forms to automorphic
forms on $\SO(d,d,\Real)/\SO(d)\times \SO(d)$,
\begin{equation}
\Phi \longrightarrow \int_{\F(\Sl(2,\Zint))} \frac{d\tau
d\bar\tau}{\tau_2^2} Z_{d,d}(\tau;g,B) ~\Phi  (\tau,\bar\tau)\,,
\end{equation}
where the ``correspondence'' $Z_{d,d}$ of~(\ref{lag}) is invariant
under $\Sl(2,\Zint)\times \SO(d,d,\Zint)$.

\paragraph{The exact $R^4$ couplings.}

Our goal in this paper is to generalize these considerations to
the eleven-dimensional membrane, and derive the exact
non-perturbative $R^4$ couplings for M-theory compactified on a
torus $\mathbb{T}^{d+1}$. Using arguments from supersymmetry and
U-duality, it was indeed shown
in~\cite{Green:1997tv,Kiritsis:1997em,Green:1997di} that the
complete non-perturbative $R^4$ coupling could be written in
terms of Eisenstein series of the U-duality group
$\exc$~\cite{Obers:1999um}
\begin{equation}
\label{cr4} f_{R^4} = \eis{\exc(\Zint)}{\irrep{string}}{s=3/2}+
\eis{\exc(\Zint)}{\irrep{membrane}}{s=1}\,.
\end{equation}
The Eisenstein series appearing in the r.h.s.\footnote{The two
terms correspond to different kinematic structures
$(t_8t_8\pm\epsilon_8\epsilon_8)R^4$, and become equal for
$d>2$~\cite{Kiritsis:1997em}. We work in units of Planck length,
so that $V_{d+1}/l_M^9=1$.} are eigenmodes of the U-duality
invariant laplacian~\cite{Kiritsis:1997em,Obers:1999um}, as
required by
supersymmetry~\cite{Green:1997di,Pioline:1998mn,Green:1999se}.
The tree-level, one-loop and D-instanton terms in~(\ref{pertexp})
arise upon expanding the Eisenstein series at weak coupling. For
our present purposes, it is however more appropriate to consider
the expansion in the limit of large volume, which commutes with
the  eleven-dimensional Lorentz group. Using the
techniques in~\cite{Obers:1999um}, we obtain
\begin{eqnarray}
f_{R^4}&=&  \frac{\pi^2 l_M^6}{3}+ \sum_{m^i\in
\Zint^{d+1}}\frac{l_M^9}{[(m^i)^2]^{3/2}}+
\pi\sum_{m^3\neq 0}\frac{l_M^9}{\sqrt{(m^3)^2}}+\nonumber\\
&&+\,\pi l_M^6 \sum_{m^3\neq 0} \left[
\frac{l_M^6}{(m^3)^2}\right]^{1/2} \mu(m^3) \exp\left( -
\frac{2\pi}{l_M^3} \sqrt{(m^3)^2} + 2\pi i m^3
C_3\right)\times\nonumber\\
&&\hphantom{+\,\pi l_M^6 \sum_{m^3\neq 0}}\!\times \left( 1 +
\mathcal{O}\left(\frac{1}{l_M^3}\right) \right) \label{memins}
\end{eqnarray}
which reveals a sum of ``perturbative terms'' and membrane
instantons~\cite{Becker:1995kb,Harvey:1999as}, wrapping subtori
$\mathbb{T}^3\subset \mathbb{T}^{d+1}$ with wrapping number
$m^{ijk}:=m^3$ (similarly, $C_3$ denotes the 3-form gauge field
$C_{ijk}$.) The instanton summation following from the previous
computation is the number theoretic function
\begin{equation}\label{mesmu}
\mu\left(m^{ijk}\right)=\sum_{n|m^{ijk}} n\, .
\end{equation}
It is therefore uniquely fixed by U-duality invariance.

Although the four-graviton amplitude is known exactly, the
derivation we have outlined is very indirect, and tells little
about the underlying fundamental degrees of freedom of M-theory.
By contrast, a first-principle derivation, would give support to
the fundamental nature of the purported degrees of freedom, as
well as important practical insights into their quantization. In
this paper, we test the proposal of  the eleven-dimensional
membrane as the elementary excitation of M-theory. In particular,
we would like to rederive the summation measure~(\ref{mesmu}) for
membrane instantons, together with the perturbative terms
in~(\ref{memins}).

\paragraph{The BPS eleven-dimensional membrane.}

At this stage, given our lack of understanding of the fundamental
degrees of freedom of M-theory, we need an act of faith. In this
paper, we will contend that at least for the purposes of deriving
this BPS amplitude,
\begin{list}{}{\setlength{\labelwidth}{10pt}}

\item[(i)] \emph{the eleven-dimensional membrane provides
the relevant degrees of freedom}. The membrane degrees of freedom
are certainly needed, since the large volume expansion of the
second term (and the first term for $d+1>3$) in~(\ref{cr4})
exhibit a sum over membrane instantons. For $d+1\geq 6$, there
are also 5-brane instantons, and therefore our computation will
certainly not give the complete result there. Finally, the first
term (for $d+1<3$) was reproduced
in~\cite{Green:1997as,Green:1999by} from a supergravity
computation, and the second term for $d=1$ in~\cite{deWit:2000ir},
but in both cases the divergence had to be fixed by hand. One may
hope that a hypothetical modular invariance of the membrane might
render the amplitude manifestly finite;

\item[(ii)] \emph{the only topology contributing
is the torus $\mathbb{T}^3$}. This assumption is based on the
fact that the only membrane instantons appearing in the
expansion~(\ref{memins}) are subtori $\mathbb{T}^3\subset
\mathbb{T}^d$. In particular, the amplitude should have
$\Sl(3,\Zint)$ modular invariance on the membrane world-volume;

\item[(iii)] \emph{all quantum fluctuations cancel, leaving
only the contribution from the zero-modes}. This is in analogy with
the string theory computation.
\end{list}
Points (ii) and (iii) are supported by a light-cone treatment of
the four-graviton amplitude using the membrane vertex operators
of~\cite{Dasgupta:2000df}. In a hamiltonian formalism the
$\mathbb{T}^3=S^1\times \mathbb{T}^2$ topology corresponds to a
one-loop closed supermembrane amplitude given by~\cite{jan}
\begin{equation}
\mathcal{A}_4 =
\STr\left(V^1\,\Delta\,V^2\,\Delta\,V^3\,\Delta\,V^4\,\Delta\right).
\label{A4}
\end{equation}
Here the $V^i$ denote the graviton vertex operators separated by
the propagator $\Delta=\int_0^\infty dt\, \exp[-t\,{H}]$ and the
supertrace $\STr$ is over the Hilbert space of ${H}$. As we are
dealing with a compactified theory the membrane spectrum becomes
discrete in a semiclassical quantization of the non-zero winding
sector~\cite{Russo}. By expanding around the classical winding
configurations the hamiltonian splits into ${H}=
H_{\mathrm{class}}+ {H}_0 + {H}_{\mathrm{int}}$: the bosonic zero
mode piece $H_{\mathrm{class}}$, a superharmonic oscillator of
the quantum fluctuations ${H}_0$ with masses given by the
windings, along with an interaction term ${H}_{\mathrm{int}}$.
Therefore the Hilbert space is spanned by the discrete and
continuous bosonic zero modes $|x^i\rangle$, the fermionic zero
modes $|\mathcal{N}\rangle$ comprising the $\mathbf{44}+
\mathbf{84}$ bosonic and $\mathbf{128}$ fermionic transverse
states of the massless $D=11$ supermultiplet, along with the
discrete eigenstates $\|m\rangle\rangle$ of
${H}_0+{H}_{\mathrm{int}}$. As the fermion zero modes $\theta_0$
do not appear in the hamiltonian and sixteen insertions of
$\theta_0$ in the trace over $|\mathcal{N}\rangle$ are needed for
a non-vanishing result, only the terms in $V^i$ of highest
fermionic degree (being four) enter the amplitude. The trace
in~(\ref{A4}) then factorizes as
\begin{eqnarray}
\mathcal{A}_4 &=&\int_0^\infty dt\,\int d^{11}x\, \langle
x|\,\,e^{-t\,H_{\mathrm{class}}} \,|x\rangle \times
\sum_{\mathcal{N}}\, \langle \mathcal{N}|(-)^{F}\,
V^1|_{\theta_0{}^4}\, V^2|_{ \theta_0{}^4}\, V^3|_{
\theta_0{}^4}\,
V^4|_{\theta_0{}^4}\,  |\mathcal{N}\rangle\times \nonumber\\
&&\hphantom{\int_0^\infty}\! \times \sum_m \langle\langle
m\|(-)^F\,e^{-t\, ({H}_0+{H}_{\mathrm{int}})} \,
\|m\rangle\rangle\, . \label{A42}
\end{eqnarray}
The trace over the fermionic zero modes yields the appropriate
tensor structure of the $R^4$ term~\cite{deRoo,pierre}
\begin{eqnarray}
\STr\left( V^1|_{\theta_0{}^4}\, V^2|_{ \theta_0{}^4}\, V^3|_{
\theta_0{}^4}\,
V^4|_{\theta_0{}^4}\right)_{|\mathcal{N}\rangle}&=&
\varepsilon^{\alpha_1 \cdots \alpha_{16}} \, \Gamma^{j_1
j_2}_{\alpha_1 \alpha_2} \cdots \Gamma^{j_{15}
j_{16}}_{\alpha_{15} \alpha_{16}}\, R_{j_1 j_2 j_3
j_4} \cdots R_{j_{13} j_{14} j_{15} j_{16}}\nonumber\\
&=& \mathcal{C}^{pqrs}\, \mathcal{C}_{pq}{}^{tu}\,
\mathcal{C}_{rt}{}^{vw}\, \mathcal{C}_{suvw}-\nonumber\\ &&-\,4\,
\mathcal{C}^{pqrs}\, \mathcal{C}_p{}^t{}_r{}^u\,
\mathcal{C}_t{}^v{}_q{}^w\, \mathcal{C}_{uvsw} +
\mbox{Ricci\hbox{-}terms}\,, \label{achtzehn}
\end{eqnarray}
where $\Gamma^{j}_{\alpha_1\alpha_2}$ denote the transverse $\SO(9)$
Dirac matrices and $\mathcal{C}^{pqrs}$ ($p,q,\dots=0,\dots 10$) is
the covariant Weyl tensor.\footnote{ This $R^4$ coupling based on
$\SO(9)$ Dirac matrices is indeed equal to the $\SO(8)$ based
$t_8\,t_8\,R^4$ coupling~\cite{deser} in the Weyl sector, they differ
however by $R_{pq}$ and $R$ terms, a fact also noted
in~\cite{pierre}.} Due to supersymmetry and the discreteness of the
spectrum of ${H}_0+{H}_{\mbox{\tiny int}}$, the (non zero-mode) last
term in~(\ref{A42}) is nothing but the Witten index associated with
this Hamiltonian. Closer inspection reveals that this index equals one
for all configurations with non-degenerate winding. We are thus left
to compute the \emph{classical} partition function $\zclass$, to which
we now turn.

\paragraph{Polyakov action of the membrane.}

In order to keep the modular symmetry $\Sl(3,\Zint)$ of the
toroidal membrane manifest, let us use the Polyakov action of the
euclidean membrane as our starting point:
\begin{equation}
\label{polm}
S=\int d^3 \sigma \sqrt{\gamma} \left( g_{ij} \gamma^{ab}
\partial_a X^i\partial_b X^j - 1\right) + i \epsilon^{abc} C_{ijk}
\partial_a X^i\partial_b X^j\partial_c X^k\,.
\end{equation}
Note that the cosmological constant term $\int\sqrt{\gamma}$ is
necessary in order to ensure the classical equivalence with the
Nambu-Goto action on-shell, and therefore reproduce the correct
weight for the membrane instantons. In particular, the
determinant of the metric does not decouple as in the string
theory case. In line with our discussion above, we restrict
ourselves to classical embeddings $X^i = m^i_a \sigma^a$ and
constant worldvolume metrics $\gamma_{ab}=u\hat\gamma_{ab}$ where
$\hat\gamma_{ab}$ is the unit volume metric. We therefore propose
to consider the integral
\begin{equation}\label{cova}
\zclass=\int_0^\infty  \frac{du}{u} u^\alpha
\int_{\F_{\Sl(3,\Zint)}} d\hat\gamma
 \sum_{m^i_a\in\Zint^{3d+3}}
e^{-\pi u ( m^i_a g_{ij} \hat\gamma^{ab} m^j_b) + \pi u^3
+2\pi i C_{ijk} \epsilon^{abc} m^i_a m^j_b m^k_c}
\end{equation}
as a candidate to reproduce the exact $R^4$ couplings
in~(\ref{cr4}) --- we shall refer to it as the \emph{covariant}
amplitude. Here $\F_{\Sl(3,\Zint)}$ denotes the fundamental
domain of $\Sl(3,\Zint)\backslash \Sl(3,\Real)/\SO(3)$, and the
integration measure $d\hat\gamma$ is the $\Sl(3,\Real)$ invariant
measure.\footnote{$\Sl(3,\Zint)$ modular forms have been
constructed in~\cite{Kiritsis:1997em}, with a different
motivation. A mathematical introduction to $\Sl(n,\Zint)$ modular
forms can be found in~\cite{Terras}.} The exponent $\alpha$ is
left unspecified at this stage, and will be fixed to $\alpha=6$
later by comparing with the light-cone amplitude. The restriction
to the fundamental domain makes the integral over $\hat\gamma$
manifestly finite, but, in sharp contrast to the string theory
case, the integral over the volume factor $u$ remains potentially
divergent and requires regularization. A natural prescription is
to rotate the integration domain of $u$ to the semi-infinite line
$u\in i\Real^+$ in the complex plane. As in the integral
representation of the Airy function ${\mbox Ai}(x)=\int_0^\infty
du \cos(u x + u^3/3)$, the oscillations render the
$u\rightarrow\infty$ integration finite, and yield the correct
Nambu-Goto euclidean saddle points. The behavior in the small
volume limit $u\rightarrow 0$ is however not controlled, and is
at the source of the difficulties in membrane theory.

For explicit computations, it is useful to parameterize the
metric through a 3-bein $\nu^a_c$ such that
$\hat\gamma^{ab}=\nu^a_c \nu^b_c$,
\begin{equation}\label{3bein}
\nu^a_b=\frac{1}{\lambda^{1/3}}\,
\pmatrix{\frac{1}{\sqrt{\tau_2}} & & \cr & \sqrt{\tau_2}& \cr & &
\lambda} \pmatrix{ 1 & \tau_1 & \omega_1 \cr & 1 & \omega_2\cr & &
1},\qquad X^i= m^i \sigma_1 + n^i \sigma_2 + p^i \sigma_3 \, .
\end{equation}
For such a configuration, the classical action reads
\begin{equation}\label{genesis}
S=\frac{\pi u }{\lambda^{2/3}} \left[ \frac{| m^i + \tau n^i +
\omega p^i|^2}{\tau_2} + \lambda^2 \left(p^i\right)^2 \right] -
\pi u^3 + 2\pi i C_{ijk} m^i n^j p^k\,,
\end{equation}
where $\tau=\tau_1 + i\tau_2$ and $\omega=\omega_1+i\tau_2\,
\omega_2$, and is integrated against the summation measure
\begin{equation}\label{mem1}
\zclass = \int_0^\infty  \frac{ du}{u}\, u^\alpha\,
\int_{\F_{\Sl(3,\Zint)}} \!\!\!\! \frac{ d\lambda~ d\tau_2~
d\tau_1~ d\omega_1~ d\omega_2} {\lambda^3 \tau_2^2}
\sum_{(m^i,n^i,p^i)\in\Zint^{3d+3}}e^{-S}\, .
\end{equation}
The precise definition of the fundamental domain
$\F_{\Sl(3,\Zint)}$ will be given below. At this stage we simply
note that the Borel subgroup of $\Sl(3,\Zint)$ allows to restrict
the integration of $(\tau_1,\omega_1,\omega_2)$ to a period
$[-1/2,1/2]$, while the integration on the volume factor $u$ is
not restricted by modular invariance.

\paragraph{From Polyakov to light-cone.}

In analogy with the string case, let us Poisson resum the
windings $m^i$ into momenta $m_i$. Changing variables from
$(u,\tau_2,\lambda)$ to $(x,y,t)$ with
\begin{equation}
u=(xyt)^{1/3}\, ,\qquad \tau_2=x^{1/2} t\, ,\qquad
\lambda=\frac{(y t)^{1/2}}{x^{1/4}}
\end{equation}
the action becomes
\begin{eqnarray}
\tilde S&=& \pi t \left[ \left(m_i+ C_{ijk} n^j p^k\right)^2 + x
\left(n^i+\omega_2 p^i\right)^2 + y \left(p^i\right)^2  -xy
\right]+\nonumber\\
&&+\, 2\pi i m_i \left(\tau_1  n^i+\omega_1 p^i\right) ,
\end{eqnarray}
to be integrated against the measure
\begin{equation}
\int_{\F_{\Sl(3,\Zint)}\times \Real^+} \frac{(xyt)^{\alpha/3} dx\
dy\ dt\ d\tau_1\ d\omega_1\ d\omega_2} {x y^2 t^3}
\sum_{m_i,n^i,p^i}e^{-\tilde S}\,.
\end{equation}
Now the integral over $(\omega_2,x,y)$ is dominated by a saddle
point at
\begin{equation}
\omega_2= - \frac{n^i p^i}{(p^i)^2}\, ,\qquad
x=\left(p^i\right)^2,\qquad y=\left(n^i-\frac{n^j p^j}{(p^j)^2}
p^i\right)^2.
\end{equation}
The saddle point action reads
\begin{equation}\label{ts}
\tilde S=\pi t \left[ \left(m_i+ C_{ijk} n^j p^k\right)^2 +
\left(n^i \right)^2 \left(p^i\right)^2 -\left(n^i p^i\right)^2
\right] + 2\pi i m_i \left(\tau_1  n^i+\omega_1 p^i\right) ,
\end{equation}
and dominates the amplitude
\begin{equation}
\zclass= \int dt\ d\tau_1\ d\omega_1~
t^{(d-8)/2+\alpha/3}\sum_{m_i, m^{ij}}
\left[(m^{ij})^2\right]^{(\alpha-6)/3} \, e^{-\tilde S} \, ,
\label{memloop}
\end{equation}
where $m^{ij}=n^i p^j-n^j p^i=\{X^i,X^j\}$, the Poisson bracket
on the $\mathbb{T}^2$ membrane worldvolume. The real part of
eq.~(\ref{ts}) is the membrane light-cone hamiltonian
\begin{equation}
H= \frac{1}{P_+}\left[ (P_i)^2+ \left\{X^i,X^j\right\}^2 \right],
\end{equation}
where $1/P_+=t$ is the Schwinger proper time. In particular, we
recognize the U-duality invariant mass spectrum
\begin{equation}\label{bpsspec}
\mathcal{M}^2= \left(m_i+ C_{ijk} m^{jk}\right)^2 +
\left(m^{ij}\right)^2 .
\end{equation}
The parameters $\tau_1$ and $\omega_1$ are Lagrange multipliers
enforcing the constraints
\begin{equation}
m_i n^i = 0\, ,\qquad m_i p^i =0\qquad  \Longrightarrow \qquad
m_im^{ij}=0
\end{equation}
which are precisely the BPS constraints. Finally, we fix the
parameter $\alpha$ by matching the measure factor to the one-loop
supergravity amplitude~\cite{Green:1997as}. The latter receives
contributions from four vertex insertions and an integral over
$(10-d)$ continuous momenta, i.e.\ $\int \frac{dt}{t}\, t^4 \,
t^{(d-10)/2}$, yielding $\alpha=6$. Happily, the factor
$[(m^{ij})]^{(\alpha-6)/3}$ in~(\ref{memloop}) drops at the same
time. Note that this derivation of the standard light-cone
membrane hamiltonian from the Polyakov action is essentially a
zero-mode version of the original argument in~\cite{DeWit:1988ig}.

\paragraph{Fundamental domain of $\Sl(3,\Zint)$.}

In order to further analyze the proposed one-loop membrane
amplitude~(\ref{mem1}), it is necessary to specify the fundamental
domain of integration. The latter can be determined as follows.
The modular group acts by right multiplication of the
3-bein~(\ref{3bein}) by $\Sl(3,\Zint)$ matrices. It is generated
by its Borel subgroup, namely the discrete Heisenberg group
\begin{equation}
\cases{ \tau_1 \longrightarrow \tau_1+ a \cr
\omega_1\longrightarrow \omega_1 + b +c\tau_1 \cr
\omega_2\longrightarrow \omega_2 + c} ,\qquad (a,b,c)\in\Zint^3
\end{equation}
and its Weyl group, namely the group of permutations of the
columns of the 3-bein. Using the Weyl group, we may thus choose a
fundamental Weyl chamber, i.e.\ to order the squared norms of the
three columns of~(\ref{3bein}) as
\begin{equation}
\frac{1}{\tau_2}< \frac{\tau_1^2+\tau_2^2}{\tau_2} <
\frac{\omega_1^2+\tau_2^2\omega_2^2 }{\tau_2}+\lambda^2
\end{equation}
and restrict the Borel moduli to
\begin{equation}
-\frac{1}{2}<(\tau_1,\omega_1,\omega_2)< \frac{1}{2}\, .
\end{equation}
In other words, $\tau=\tau_1+i\tau_2$ is restricted to the
standard fundamental domain of $\Sl(2,\Zint)$, $\omega_1+\tau
\omega_2$ is restricted to the torus of complex modulus $\tau$,
while the extra modulus $\lambda$ takes values on the half-line
\begin{equation}
\lambda^2 \tau_2 > \tau_1^2+\tau_2^2(1-\omega_2)^2-\omega_1^2\,.
\end{equation}
The volume modulus $u$, on the other hand, should be integrated
from 0 to $\infty$.

\paragraph{Membrane instantons.}

Let us now consider the large volume expansion of~(\ref{mem1}), or
equivalently the Fourier expansion in the periodic modulus
$C_{ijk}$. In analogy with the one-loop string amplitude, one
should decompose the summation over the integers $(m^i,n^i,p^i)$
into their orbits under $\Sl(3,\Zint)$, and unfold the
integration region accordingly. Orbits of  $\Sl(3,\Zint)$ are
classified by the rank of the matrix $(m^i,n^i,p^i)$, hence we
have now four different contributions. The simplest case to
consider is the non-degenerate orbit of rank 3: the fundamental
domain can be unfolded to the full ``upper half plane''
\begin{equation}
(\tau_2,\lambda,u)\in\Real^+\, ,\qquad
(\tau_1,\omega_1,\omega_2)\in\Real\, .
\end{equation}
The integral over the Borel parameters
$(\tau_1,\omega_1,\omega_2)$ of~(\ref{genesis}) and~(\ref{mem1})
is exactly gaussian, and gives a classical action
\begin{eqnarray}
\Re(S)&=&\frac{\pi u}{\lambda^{2/3} \tau_2
(m^{ij})^2\left(p^i\right)^2}
\Big\{\left(m^{ijk}\right)^2\left(p^i\right)^2+\tau_2 \lambda^2
\left(m^{ij}\right)^2\left[(p^i)^2\right]^2+\nonumber\\
&&\hphantom{\frac{\pi u}{\lambda^{2/3} \tau_2 (m^{ij})^2(p^i)^2}
\Big\{}\!+\tau_2^2 \left[(m^{ij})^2\right]^2\Big\} - \pi u^3
\end{eqnarray}
supplemented by the same coupling $2\pi i C_{ijk} m^i n^j p^k$ to
the background 3-form. Here $m^{ij}=n^{[i} p^{j]}$ and
$m^{ijk}=m^{[i} n^j p^{k]}$. In terms of the $(x,y,t)$ variables
this is
\begin{equation}
S=\pi\left\{
\frac{(m^{ijk})^2}{t(m^{ij})^2}+\frac{(m^{ij})^2}{(p^i)^2}xt
+\left(p^i\right)^2 yt\right\} - \pi xyt  +2\pi i C_{ijk}
m^{ijk}\,.
\end{equation}
Finally, we can integrate over $(x,y,t)$, and obtain in the
saddle point approximation
\begin{equation}\label{memnd}
A_\mathrm{nondeg}=\sum_{m^{ijk}} \mu'\left(m^{ijk}\right)
\frac{e^{-2\pi \sqrt{(m^{ijk})^2}+ 2\pi i C_{ijk} m^{ijk}}}
{\sqrt{(m^{ijk})^2}}\,.
\end{equation}
The summation measure $m^{ijk}$ can be computed by noting that
the sum has to go over $\Sl(3,\Zint)$ orbits of the integer matrix
$(m^i,n^i,p^i)$: a representative can be chosen~as
\begin{equation}
\pmatrix{ m^i\cr n^i\cr p^i}= \pmatrix{m &j &k & m^4 & m^5 &\cdots
\cr 0 & n & l &n^4 & n^5 &\cdots \cr 0 & 0 & p & p^4 & p^5
&\cdots}
\end{equation}
with $m,n,p>0$ and $0\leq j<n, 0\leq k,l<p$. Hence the summation
measure is given by the number-theoretic function
\begin{equation}\label{mesm}
\mu'\left(m^{ijk}\right)= \sum_{n|m^{ijk}}\sum_{p|(m^{ijk}/n)} n
p^2\,.
\end{equation}
\looseness=1 Importantly, this geometric summation measure
disagrees with the summation measure  predicted by U-duality
in~(\ref{mem1}). In addition to these instanton contributions,
the amplitude also contains perturbative terms corresponding to
the degenerate orbits of rank 0,1,2, which should reproduce the
first three terms in~(\ref{memins}), respectively. In view of the
discrepancy already observed at the level of the non-degenerate
orbit, we will refrain from discussing those in detail. We
however observe that, had our semi-classical proposal succeeded
in reproducing the membrane instantons, it would also have
produced the correct perturbative terms, since those are related
by U-duality.

\paragraph{Intermezzo.}

Before proceeding, some comments are in order.
\begin{list}{}{\setlength{\labelwidth}{10pt}}
\item[(i)] the covariant amplitude~(\ref{cova}) reproduces
both the correct BPS mass spectrum~(\ref{bpsspec}) and classical
saddle point values~(\ref{memnd}) for membrane instantons. This
comes as no surprise, given the equivalence between the Polyakov,
light-cone and Nambu-Goto formulations of the membrane at the
classical level.

\item[(ii)] one advantage of the covariant amplitude is its
manifest modular invariance under $\Sl(3,\Zint)$, which is hidden
in the standard light-cone approach. One should therefore
restrict to the fundamental domain of $\Sl(3,\Zint)$ in order to
avoid multiple countings; in contrast to the string theory case,
modular invariance however does not imply the finiteness of
membrane theory yet, since the integration on the volume $u$
remains unbounded.

\item[(iii)] while the classical saddle points are correctly
reproduced, the covariant amplitude predicts the wrong instanton
summation measure; in particular, it is not U-duality invariant
--- in fact, this is easily seen in the $d=2$ case with $C=0$.
Although the classical Bianchi identities and equations of motion
following from the Polyakov action for the membrane do fall into
U-duality multiplets~\cite{Duff:1990hn}, this is not sufficient
to ensure the invariance of the classical partition function. In
particular, U-duality requires the exchange of momentum $P_i$ and
the composite winding number $\{X^i,X^j\}$, under which the
measure is not manifestly invariant.

\item[(iv)] in order to generate the proper $R^4$ amplitude, the
membrane partition function should be an eigenmode of a
combination of the $\Sl(3,\Real)/\SO(3)$ and $\exc$ laplacian.
This is not the case, even though the values of the saddle points
are indeed eigenmodes of the $\exc$ laplacian.

\item[(v)] the discrepancy between the geometric summation
measure~(\ref{mesm}) and the one predicted by U-duality
in~(\ref{mem1}) implies that the contribution of the non-zero
modes to the covariant amplitude is not unity as a naive extrapolation from
the string case might have suggested, but instead should be given
by the ratio $\mu(N)/\mu'(N)$ where $N=m^{ijk}$. This is an
interesting prediction about the interacting world-volume theory
of the membrane that would be interesting to~verify.
\end{list}
Keeping with our act of faith, we now propose that the
shortcomings of our proposal~(\ref{cova}) can be repaired by
enforcing modular invariance and U-duality explicitly. More
precisely, our task is \emph{to construct a partition function
$\Xi_{d+1}(\gamma_{ab};g_{ij},C_{ijk})$ for mappings of
$\mathbb{T}^3$ into $\mathbb{T}^{d+1}$, invariant under
$\Sl(3,\Zint)\times \exc{(\Zint)}$}, such that, upon integration
on the fundamental domain of $\Sl(3,\Zint)$,\footnote{Here we
include the integration on the volume factor $\det(\gamma)$ in
$\Xi_{d+1}$.} we reproduce the non-perturbative $R^4$ couplings:
\begin{equation}
\int_{\F(\Sl(3,\Zint))} \Xi_{d+1} =
\eis{\exc(\Zint)}{\irrep{string}}{s=3/2}+
\eis{\exc(\Zint)}{\irrep{membrane}}{s=1}\,.
\end{equation}
More generally, integrating $\Xi_{d+1}$ against an $\Sl(3,\Zint)$
automorphic form should produce a lift to a $\exc(\Zint)$ modular
form. This is a standard problem in the theory of automorphic
forms, that goes under the name of ``theta'' correspondence.

\paragraph{Theta correspondences.}

In order to gain further insight into this problem, it is useful
to return to the one-loop string amplitude~(\ref{1loop}) once
again. The invariance under $\Sl(2,\Zint)\times \SO(d,d,\Zint)$
of the partition function $Z_{d,d}(\tau;g,B)$  of
the string zero-modes is a rather subtle phenomenon: only the
modular invariance $\Sl(2,\Zint)$ is manifest in the lagrangian
picture~(\ref{lag}), while the $\SO(d,d,\Zint)$ symmetry becomes
apparent in the hamiltonian picture~(\ref{ham}), at the cost of
losing the manifest modular symmetry. Indeed, the partition
function~(\ref{lag}) is a special case of a theta series
\begin{equation}
\theta_{\Sp(g)}(\Omega_{AB}) = \sum_{m^A\in \Zint^{g}} e^{-\pi m^A
\Omega_{AB} m^B}
\end{equation}
invariant under the symplectic group $\Sp(g,\Zint)$ acting by
fractional linear transformations on the period matrix $\Omega\in
\Sp(g,\Real)/\U(g)$, through Poisson resummation on the integers
$m^A$. In fact,
\begin{equation}
Z_{d,d}(\tau;g,B)=\theta_{\Sp(2d)}(\tau \otimes (g+b))\,,
\end{equation}
where the tensor product provides an embedding $\Sl(2)\times
\SO(d,d)\subset \Sp(2d)$. The pair $\Sl(2)\times \SO(d,d)$, such
that the centralizer of either group is equal to the other, is
known as a dual pair~\cite{Howe}. In fact, there exists a
discrete set of elements in $\Sp(2d)$ preserving the
decomposition $\Sl(2)\times \SO(d,d)\subset \Sp(2d)$, so that the
partition function $Z_{d,d}$ is actually invariant under a bigger
group mixing target-space and world-sheet~\cite{Rabino}. If our
proposal holds, we would therefore also have a symmetry mixing
target-space and world-volume in membrane theory.

Our task is therefore to find a generalization of
$\theta_{\Sp(g)}$, such that the restriction to an
$\Sl(3,\Zint)\times\exc(\Zint)$ subgroup will provide our
$\Xi_{d+1}$. Remarkably, such problems are the subject of much
interest in the theory of automorphic forms, and many results are
available in the mathematical literature. The relevant results
for us are the following:
\begin{list}{}{\setlength{\labelwidth}{10pt}}
\item[(i)] A classification of dual pairs $G_1 \times G_2 \subset G$
exists~\cite{Dynkin,Rubenthaler}. The part of it relevant for our
purposes is reproduced in table~\ref{tab:1}. For the problem at
hand, we are interested in the following pairs,
\begin{eqnarray}
\Real^+\times \Sl(3)\times \Sl(2) &\subset& \Sl(5)\,, \nonumber\\
\Real^+\times \Sl(3)\times \Sl(2) \times \Sl(3)&\subset& \Sl(3)^3
\subset E_{6(6)}\,.
\nonumber\\
\Real^+\times \Sl(3)\times \Sl(5) &\subset& \Sl(8)\,,\nonumber \\
\Real^+\times \Sl(3)\times \SO(5,5) &\subset&\Sl(4)\times
\SO(5,5)\subset E_{8(8)}\,,
\nonumber\\
\Sl(3) \times E_{6(6)} &\subset& E_{8(8)} \label{excor}
\end{eqnarray}
which allow to lift $\Sl(3,\Zint)$ modular forms to $\exc(\Zint)$
for $d=1,2,3,4,5$, respectively. We have included the $d=5$
example for completeness, although for M-theory on
$\mathbb{T}^6$, due to M5-brane instantons, one should not expect
that the membrane provide all the relevant degrees of freedom.
This last example is nevertheless quite attractive, 
since it contains all the other ones, and involves no volume
factor  at all: the integral becomes manifestly finite when
restricting to the fundamental domain of $\Sl(3,\Zint)$. The
other examples listed in table~\ref{tab:1} might have some
relevance in other M-theoretical contexts.

\TABLE[t]{\begin{tabular}{|c@{\hspace{-0.1pt}}|c|c
@{\hspace{-0.1pt}}|}\hline $G$ & $n$ & correspondences\\ \hline
$E_8$ & 29 & $E_7\times \Sl(2),\ E_6\times \Sl(3),\
\SO(5,5)\times \Sl(4),\
\Sl(3)\times \Sl(2),\ F_4\times G_2$ \\
$E_7$ & 17 & $ \SO(6,6)\times \Sl(2),\ \Sl(6)\times \Sl(3),\
G_2\times \Sl(2),\ G_2 \times \Sp(6),\ F_4\times \Sl(2)$\\
$E_6$ & 11 & $\Sl(3)\times \Sl(3)\times \Sl(3),\ \Sl(6)\times
\Sl(2),\
G_2 \times \Sl(3)$\\
$\SO(5,5)$ & 7 & $\Sl(2)\times \Sl(2)\times \Sl(4),\
\Sp(4)\times \Sp(4)$\\
$\Sl(8)$ & 7 & $\Real^+ \times \Sl(3) \times \Sl(5)$\\
$\Sl(5)$ & 4 & $\Real^+\times \Sl(2)\times \Sl(3)$\\
$F_4$& 8& $G_2\times \Sl(2),\ \Sl(3)\times \Sl(3),\
\Sp(6)\times \Sl(2)$\\
$G_2$& 3& $\Sl(2)\times \Sl(2)$\\
\hline
\end{tabular}%
\caption{Non-exhaustive list of correspondences for exceptional
groups, and dimension $n$ of the ``smallest'' representation. All
groups are assumed in the split real form.}\label{tab:1} }

\item[(ii)] For any simply-laced split group
$G$~\cite{KazhdanS,Kazhdan,grs2} (and for some non simply-laced
groups, see e.g.~\cite{grs}), there exists an essentially unique
minimal unitary representation $\pi$ of $G$, on a Hilbert space
$H$ of functions of $n$ variables~\cite{Joseph}. This is
analogous to the Weyl representation of the symplectic group
$\Sp(g)$ on functions of $g$ variables~\cite{tata}, where the
generators of a maximal Heisenberg subgroup are represented by
the translations and multiplications by characters of the
Schr\"odinger representation, and the action of the Heisenberg
subgroup is extended to that of the full group $G$ by a Weyl
reflection represented by the Fourier transform with respect to
the $n$ variables. These minimal representations, and in
particular the 29-dimensional representation of $E_{8(8)}$ and
the associated modular forms, remain to be constructed
explicitly. A first step in this direction was taken
in~\cite{Gunaydin:2000xr} with the explicit construction of a
quasi-conformal (non-unitary) representation of $E_{8(8)}$ on
$\Real^{57}$. The dimension of the minimal representation was
first derived in~\cite{Joseph}, and is shown in table~\ref{tab:1}
for various groups of interest. In contrast to the symplectic
case, the representation of the Heisenberg subalgebra requires
some cubic characters, and the closure of the algebra implies some
important identities for their Fourier transform.

\item[(iii)] There exists a vector $f\in H$ invariant under
the maximal compact subgroup $K\subset G$; and a distribution
$\delta$ in $H^*$ invariant under a discrete subgroup $G(\Zint)$
generated by the integer translations and Weyl reflections. From
these two objects one can construct an automorphic form $\theta$
on $G(\Zint)\backslash G/K$, given by the matrix element
$\theta(\Omega)=\langle \delta, \pi(\Omega) f \rangle$. This
generalizes the theta function for the symplectic case, where
$f(x)= e^{-\pi (x^A)^2}$, and
$\delta(x)=\sum_{m\in\Zint^g}\delta(x-m)$. In general, one
expects $\delta$ to involve a sum on a $n$-dimensional lattice,
where $n$ is the dimension in which the minimal representation is
realized. The invariance of $\theta$ under Weyl reflection
requires a Poisson resummation, which follows from the Fourier
transform identity for the non-gaussian character found in (ii).
\end{list}
Unfortunately, although theta correspondences are known to exist,
they remain elusive objects. It would therefore be very desirable
to construct explicit formulas analogous to~(\ref{lag}). From the
mathematical point of view, this would provide a wealth of new
Poisson resummation formula for non-gaussian
characters\footnote{In fact cubic, since the membrane couples to
the $C$-field through a cubic coupling, while the Polyakov action
becomes also cubic when the volume $u$ is eliminated.}
generalizing the ones found in~\cite{etingof}, and may have
interesting applications in number theory. For our purposes, it
would be very interesting to compare them to our proposal
in~(\ref{mem1}) in order to extract the quantum measure for the
membrane theory. An exciting possibility is that the
$\Sl(3)\times E_{6(6)}\subset E_{8(8)}$ exceptional theta
correspondence in~(\ref{excor}) would reproduce the full $R^4$
couplings: in addition to curing the small volume divergence of
the membrane, it would also imply that BPS membranes contain
five-branes. This we believe would be an important hint that
membranes may indeed be the fundamental degrees of freedom of
M-theory.

\acknowledgments

The authors are grateful to P.~Etingov, E.~Kiritsis, N.~Obers,
K.~Peeters, A.~Polishchuk, J.~Russo, P.~Vanhove and N.~Wyllard
for useful discussions, to R.~Borcherds and E.~Rabinovici for
directing them to the right door, and especially to D.~Kazhdan
for enlightening discussions and useful guidance in the math
literature. B.P. and A.W. express gratitude to AEI for
hospitality during the beginning of this project. Research
supported in part by NSF grant PHY99-73935, the David and Lucile
Packard foundation, and the European network HPRN-CT-2000-00131.

\newcommand{\mathAG}[1]{\href{http://xxx.lanl.gov/abs/math.AG/#1}{\tt
math.AG/#1}}

\end{document}